\documentclass[journal=langd5,manuscript=article]{achemso}

\usepackage[version=3]{mhchem} 
\usepackage[dvips]{color}

\author{Francisco J. Montes Ruiz-Cabello}
\affiliation[University of Granada]{Biocolloid and Fluid Physics Group, Applied Physics Department, Faculty of 
Sciences, University of Granada, E-18071 Granada (Spain)}
\author{Halim Kusumaatmaja}
\affiliation[Oxford University]{The Rudolf Peierls Centre for Theoretical Physics, Oxford University, 1 Keble Road, 
Oxford OX1  
3NP, U.K.}
\author{Miguel A. Rodr\'{i}guez-Valverde}
\email{marodri@ugr.es}
\affiliation[Corresponding author. Tel.: +34 958 24 00 25; fax: +34 958 24 32 14.]{Biocolloid and Fluid Physics 
Group, Applied Physics Department, Faculty of Sciences, University of Granada, E-18071 Granada (Spain)}
\author{Julia Yeomans}
\affiliation[Oxford University]{The Rudolf Peierls Centre for Theoretical Physics, Oxford University, 1 Keble  Road, 
Oxford OX1  
3NP, U.K.}
\author{Miguel A. Cabrerizo-Vílchez}
\affiliation[University of Granada]{Biocolloid and Fluid Physics Group, Applied Physics Department, Faculty of 
Sciences, University of Granada, E-18071 Granada (Spain)}

\title{Modelling the corrugation of the three-phase contact line perpendicular to a chemically striped substrate}

\begin{document}
\begin{abstract}
We model an infinitely long liquid bridge confined between two plates chemically patterned by stripes of same width 
and different contact angle, where the three-phase contact line runs, on average, perpendicular to the stripes. This 
allows us to study the corrugation of a contact line in the absence of pinning. We find that, if the spacing between 
the plates is large compared to the length scale of the surface patterning, the cosine of the macroscopic contact 
angle corresponds to an average of cosines of the intrinsic angles of the stripes, as predicted by the Cassie 
equation. If, however, the spacing becomes of order the length scale of the pattern there is a sharp crossover to a 
regime where the macroscopic contact angle varies between the intrinsic contact angle of each stripe, as predicted by 
the local Young equation. 
The results are obtained using two numerical methods, Lattice Boltzmann (a diffuse interface approach) and Surface 
Evolver (a sharp interface approach), thus giving a direct comparison of two popular numerical approaches to 
calculating drop shapes when applied to a non-trivial contact line problem. We find that the two methods give 
consistent results if we take into account a line tension in the free energy. In the lattice Boltzmann approach, the 
line tension arises from discretisation effects at the diffuse three phase contact line.

 \end{abstract}


\section{Introduction}\label{introduction}

The contact angle, between the tangent to a drop and the solid surface that supports it, is an important concept in 
many different applications such as coatings, detergency, printing, adhesives and dentistry \cite{Good}. The contact 
angle provides information about how a liquid spreads on a surface in a given solid--liquid--gas system and it allows 
an estimation of the surface energy of solids \cite{Tavana}. A drop of liquid on an ideal solid surface, neglecting 
gravity effects, will be a spherical cap with a circular contact line between the three phases and the same contact 
angle at all points around the contact line. However, topographic and chemical defects on the surface can lead to a 
contact line that is not circular, and a contact angle that varies along the contact line \cite{Valverde1}. 
Traditionally such a variation, has usually been ignored as, for drops much larger than any surface feature, an 
effective, average contact angle is measured regardless of the observation direction. However, now that it is 
relatively easy to design well-defined micropatterned surfaces, and observe the behaviour of drops with dimensions of 
order the surface patterning \cite{GauLipowsky,Vrancken,Youngblood,Quere,Darhuber,Stone,Fermigier}, variations in 
contact angle around the drop can be substantial and can be measured.

\begin{figure} 
\begin{center}
\includegraphics[width=\columnwidth,angle=0]{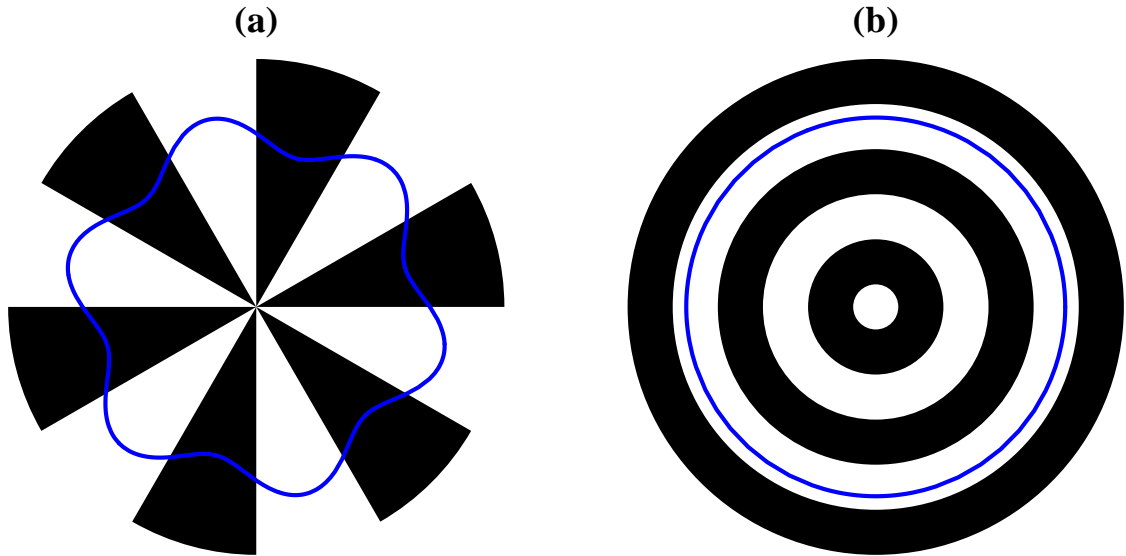}
\caption{A chemically--patterned substrate that will lead to a three-phase contact line that is (a) corrugated, but 
not  
pinned, (b) pinned, but not corrugated.}
\label{contact}
\end{center}
\end{figure}

We distinguish between two very distinct behaviours of a three--phase contact line on a patterned surface which can 
affect the uniqueness of the contact angle \cite{JohnsonDettre}. If the boundaries between regions of different 
wettability are perpendicular to the contact line as, for example, in Figure \ref{contact}(a), then the interface 
will adopt a corrugated shape to minimise its  free energy and the contact angle will vary along the contact line. 
Such variation is known as \emph{contact angle multiplicity}. This configuration corresponds to thermodynamic 
equilibrium and the final drop shape is independent of the initial conditions \cite{Li,Iliev,Schwartz1,Schwartz2}. 
If, however, the boundaries are parallel to the contact line, the drop can jump or be pinned leading to \emph{contact 
angle hysteresis} \cite{Schwartz1,Schwartz2,Decker1,Iwamatsu,Decker2,Kusumaatmaja3}. Now the final drop state depends 
on its dynamic history. An example of surface patterning where this behaviour will dominate is depicted in Figure 
\ref{contact}(b). A drop placed at the centre of the axial pattern will spread until it is pinned by a (relatively) 
hydrophobic circle -- which particular circle will be selected by the initial volume and energy of the drop, and will 
in turn determine the measured value of the contact angle. 

In general, on real surfaces, both multiplicity and hysteresis in contact angle will be important, and the presence 
of one is typically accompanied by the other. However, there are experimental and theoretical works in the literature 
\cite{Marmur1,Marmur2,Valverde2,Brandon} which have considered geometries aiming to separate the two effects, and 
this is a helpful way to investigate a complicated problem. We follow this approach here, concentrating on modelling 
the equilibrium state of a liquid bridge confined between two chemically striped plates such that the contact lines 
run, on average, perpendicular to the stripes. We describe the crossover between the behaviour of the contact line 
and the contact angle when the spacing between the plates is large, or small, compared to the length scale of surface 
patterning.

A second aim of our work is to compare two different numerical approaches, a diffuse interface model, solved using a 
lattice Boltzmann code \cite{Briant,KusumaatmajaEPL}, and Surface Evolver \cite{Brakke}, an algorithm which assumes a 
sharp interface. We discuss the effects of the finite thickness of the interface and compare the efficiency and 
applicability of the two algorithms.

In next section we describe the geometry of the model, outline the diffuse interface and Surface Evolver approaches 
and list the parameters used in the simulations. The results are then displayed and discussed. In particular, we 
explain how the surface patterning affects the macroscopic contact angle. Next, we summarise the paper and compare 
the two numerical methods.

\section{Methods}\label{methods}

\subsection{Geometry}
 
\begin{figure} 
\begin{center}
\includegraphics[width=\columnwidth,angle=0]{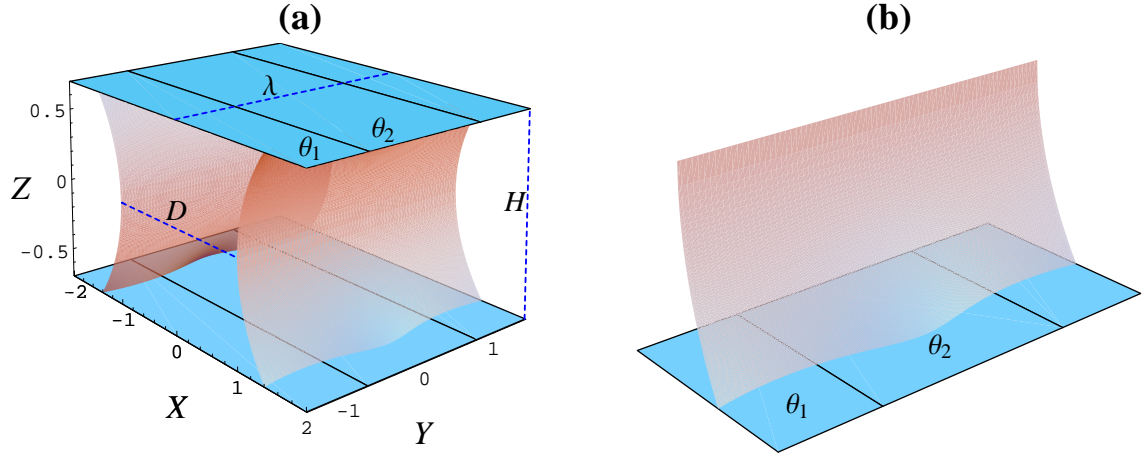}
\caption{(a) Liquid bridge between chemically--striped walls and (b) magnification of the liquid--gas interface at 
the three--phase contact line.}
\label{geometry}
\end{center}
\end{figure}

An infinitely long liquid bridge was confined between two chemically--patterned walls as shown in Figure  
\ref{geometry}. These walls were smooth planes perpendicular to the $z$-axis, lying at $z = 1$ and $z = H+1$, and 
infinite in the $y$-direction. Both walls were patterned with stripes of equal width $\left(\lambda/2\right)$, lying 
parallel to the $x$-axis. Stripes on the two walls faced each other. The intrinsic contact angles of the stripes were 
taken to alternate between  $\theta_i=60^\circ$ and  $\theta_i=30^\circ$. The width $D$ of the liquid bridge, i.e. 
the bridge dimension parallel to the stripes, was fixed sufficiently large that there was no interaction between the 
two liquid--gas interfaces $\left(D\gg\xi\right)$. For the diffuse interface model, the interface thickness $\xi$ 
provides an additional length scale. The effects of gravity are neglected in this paper.

The interface lies, on average, parallel to the $y$-axis, with an oscillation because the fluid prefers to wet the 
stripes of lower contact angle, as shown in Figure \ref{geometry}(b). Our aim was to understand how the interface 
shape varies with the interface thickness, $\xi$, the pattern period, $\lambda$, and the spacing between the plates, 
$H$. 

We next summarise the two numerical approaches that were used to calculate the shape of the liquid--gas interface. 
These methods are two widely-used numerical approaches to calculating drop shapes.

\subsection{Diffuse interface model}
 
The first numerical method is a mesoscale simulation approach where the equilibrium properties of the drop are  
modelled by a continuum free energy:
  \begin{equation} 
    \Psi = \int_V (\psi_b(n)+\frac{\kappa}{2} (\partial_{\alpha}n)^2) dV
    + \int_S \psi_s(n_s) dS . 
    \label{eq3}
    \end{equation}   
$\psi_b(n)$ is a bulk free energy term which we take to be \cite{Briant}:
    \begin{equation}
    \psi_b (n) = p_c (\nu_n+1)^2 (\nu_n^2-2\nu_n+3-2\beta\tau_w) \, ,
    \end{equation}                                      
where $\nu_n = {(n-n_c)}/{n_c}$, $\tau_w = {(T_c-T)}/{T_c}$ and $n$, $n_c$, $T$, $T_c$ and $p_c$ are the local 
density, critical density, local temperature, critical temperature and critical pressure of the fluid  
respectively. The parameter $\beta$ is related to the density contrast between the liquid and gas phases.
This choice of free energy leads to two coexisting bulk phases (liquid and gas) of density 
$n_c(1\pm\sqrt{\beta\tau_w})$. The second term in Eq.\ (\ref{eq3}) models the free energy associated with any 
interfaces in the system. The parameter $\kappa$ is related to the surface tension via $\gamma =  
{(4\sqrt{2\kappa p_c} (\beta\tau_w)^{3/2} n_c)}/3$ and the interface thickness via $\xi = \sqrt{\kappa 
n_c^2/4\beta\tau_wp_c}$ \cite{Briant}. The final term in Eq.\ (\ref{eq3}) describes  
the interactions between the 
fluid and the solid surface. Following Cahn \cite{Cahn}, the surface energy density is taken to be $\psi_s (n) =  
-\phi 
\, n_s$, where $n_s$ is the value of the fluid density at the surface. The strength of interaction, and hence  
the 
local intrinsic contact angle, $\theta_i$, is parameterized by the variable $\phi$. In our simulations,  
chemically heterogeneous surfaces are simply modelled by setting the value of $\phi$ appropriately at every  
site of the solid surface lattice \cite{Briant}.

The dynamics of the drop is described by the continuity  (\ref{eq1}) and the Navier-Stokes equations (\ref{eq2}):
\begin{eqnarray}
    &\partial_{t}n+\partial_{\alpha}(nu_{\alpha})=0 \, , 
    \label{eq1}\\
    &\partial_{t}(nu_{\alpha})+\partial_{\beta}(nu_{\alpha}u_{\beta}) = 
    - \partial_{\beta}P_{\alpha\beta}+ \nu \partial_{\beta}[n(\partial_{\beta}u_{\alpha} +  
\partial_{\alpha}u_{\beta} 
+ \delta_{\alpha\beta} \partial_{\gamma} u_{\gamma}) ] \, , 
    \label{eq2}
\end{eqnarray}
where $\mathbf{u}$, $\mathbf{P}$, and $\nu$ are the local velocity, pressure tensor, and kinematic viscosity 
respectively. The thermodynamic properties of the system appear in the equations of motion through the pressure  
tensor $\bf P$: mechanical equilibrium is equivalent to minimising the free energy. Eqs.~\ref{eq1} and  
\ref{eq2} are solved using a Lattice Boltzmann algorithm which is described in detail in \cite{Briant,Swift,Succi}.

The liquid drop was initialised as a cuboid confined in the $z$-direction by the two chemically-striped surfaces, 
with periodic boundary conditions applied in the $x$ and $y$ directions. The simulation parameters which were used 
for all the numerical calculations were: $\kappa = 0.002$, $p_c = 1/8$, $n_c = 3.5$, $T = 0.4$, $T_c = 4/7$, $\nu = 
0.1$, and $\beta=0.1$, while those specific to a particular simulation are given at the appropriate place in the 
text. 

\subsection{Surface Evolver} 

Surface Evolver is a public domain software, developed by Kenneth Brakke \cite{Brakke}, which minimises the surface 
energy of a  given volume of liquid within a prescribed geometry. The liquid--gas, liquid--solid and gas--solid 
interface  
energies, and hence implicitly any contact angles, are inputs to the model. If Surface Evolver is able to find the 
correct minimum as in \cite{Brandon,Brinkmann,Buehrle,Gea}, it provides a useful alternative to diffuse interface 
models, both because it is computationally quicker, and because in many cases a sharp interface represents the 
physically appropriate limit.

To perform the Surface Evolver simulations the liquid drop was initialised as a cuboid confined in the $z$-direction 
by the two chemically--striped surfaces. 
Symmetry demands that the interface must meet the $x-z$ plane at right-angles at the centre of each of the chemical 
stripes; we took advantage of this symmetry and imposed neutrally wetting walls at the centres of neighbouring 
stripes without altering the interface profiles.


\subsection{Measuring the contact angle and contact line corrugation}
 
Once the interface shape had been calculated, we recorded the interface profiles for different values of $y$.  
The macroscopic contact angle $\theta(y)$ was obtained by fitting a circle to the entire interface profile at $y$ and 
measuring  
its angle of intersection with $z=1$. There will be a small correction because the contact line is not parallel  
to the $y$-axis, but this was found to be negligible. This definition of contact angle is similar to that
typically used in experiments, e.g. in Axisymmetric Drop Shape Analysis (ADSA) \cite{Wege,Lam,Barbieri}. 

The distortion of the contact line, $\Delta x$, was measured as the distance between the 
maximum and the minimum values of its $x$ coordinate which occur, by symmetry, in the centre of the hydrophilic  
and hydrophobic stripes respectively. When the interface was diffuse, we defined its position as that where the  
density took the mean of its values in the liquid and gas phases.

\section{Results}
\label{results}

We aim to understand how the contact line corrugation, $\Delta x$, depends on the interface thickness, $\xi$,  
the width of the stripes, $\lambda/2$, and the height of the slab, $H$. To present the results we will scale all  
lengths to the spatial period $\lambda$.
 
\begin{figure} 
\begin{center}
\includegraphics[width=0.7\columnwidth,angle=0]{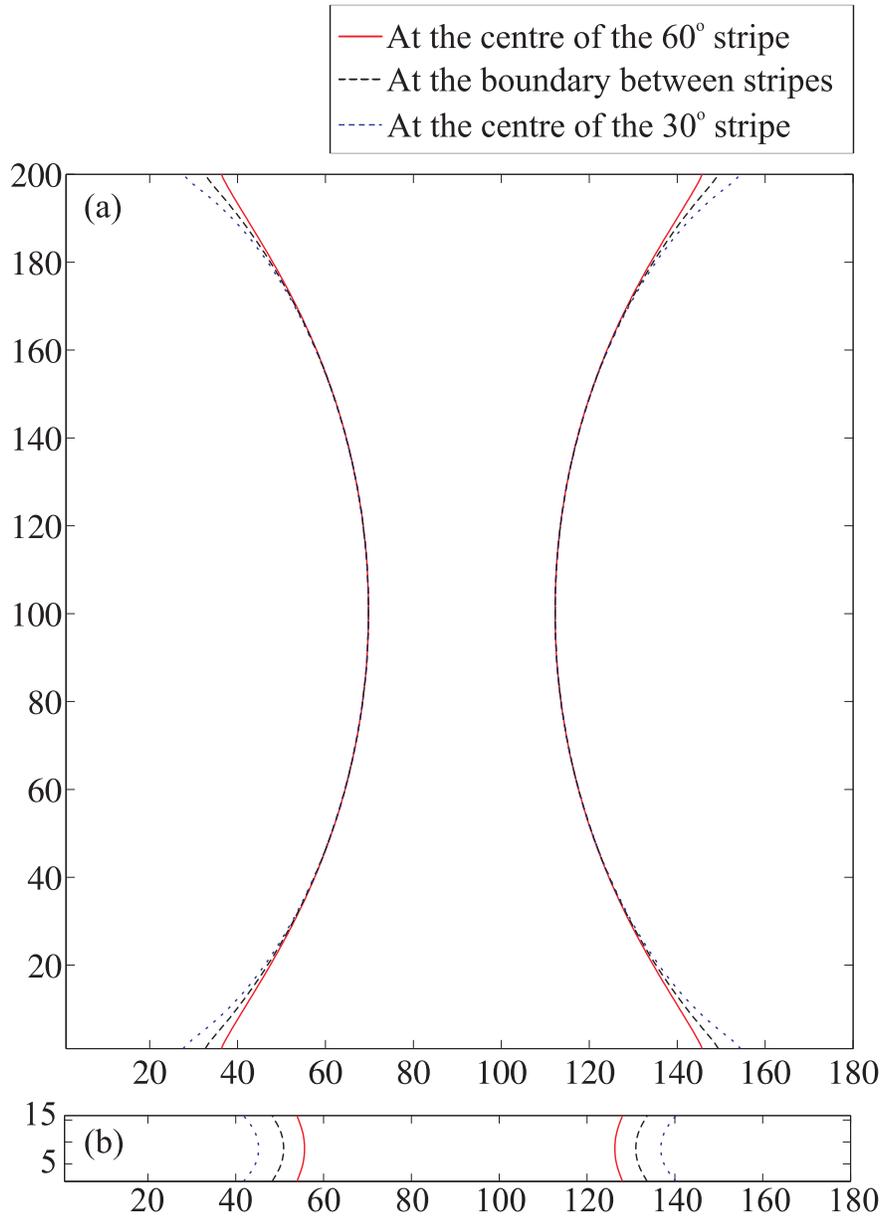}
\caption{Lattice Boltzmann results for the interface profile in the $x-z$ plane at the centre of the 
$\theta_i=60^\circ$ stripe, the centre of  
the $\theta_i=30^\circ$ stripe and the boundary between stripes for (a) $H/\lambda=200/80$, (b) $H/\lambda=16/80$, 
and $\xi/\lambda=0.016$. $H$, $\lambda$, and $\xi$ are respectively the height of the channel, the spatial period of 
the surface patterning, and the thickness of the interface.}
\label{cross-section}
\end{center}
\end{figure}

We first consider the variation of the amplitude of the contact line distortion with the distance between the  
plates. There are two distinct regimes. For large bridge heights, the magnitude of the contact line distortion 
becomes 
independent of $H/\lambda$. This occurs because the corrugations in the interface decay with height over a 
\emph{healing}
length, of order $\lambda/2\pi$ \cite{Degennes}, small compared to the spacing between the plates. Hence the 
interface away from  
the surfaces is not corrugated. This is illustrated in Figure \ref{cross-section}(a), which shows cross sections 
across  
the liquid bridge at values of $x$ corresponding to the centre of a hydrophobic stripe, the border between  
stripes and the centre of a hydrophilic stripe for $H/\lambda=200/80$. 
Figure \ref{cross-section}(b) is a similar plot, but for plate separations $H/\lambda=16/80$. Now the decay length  
of the corrugation along $z$ is larger than $H$ and it is favourable for the corrugation to persist for all $z$.

\begin{figure} 
\begin{center}
\includegraphics[width=\columnwidth,angle=0]{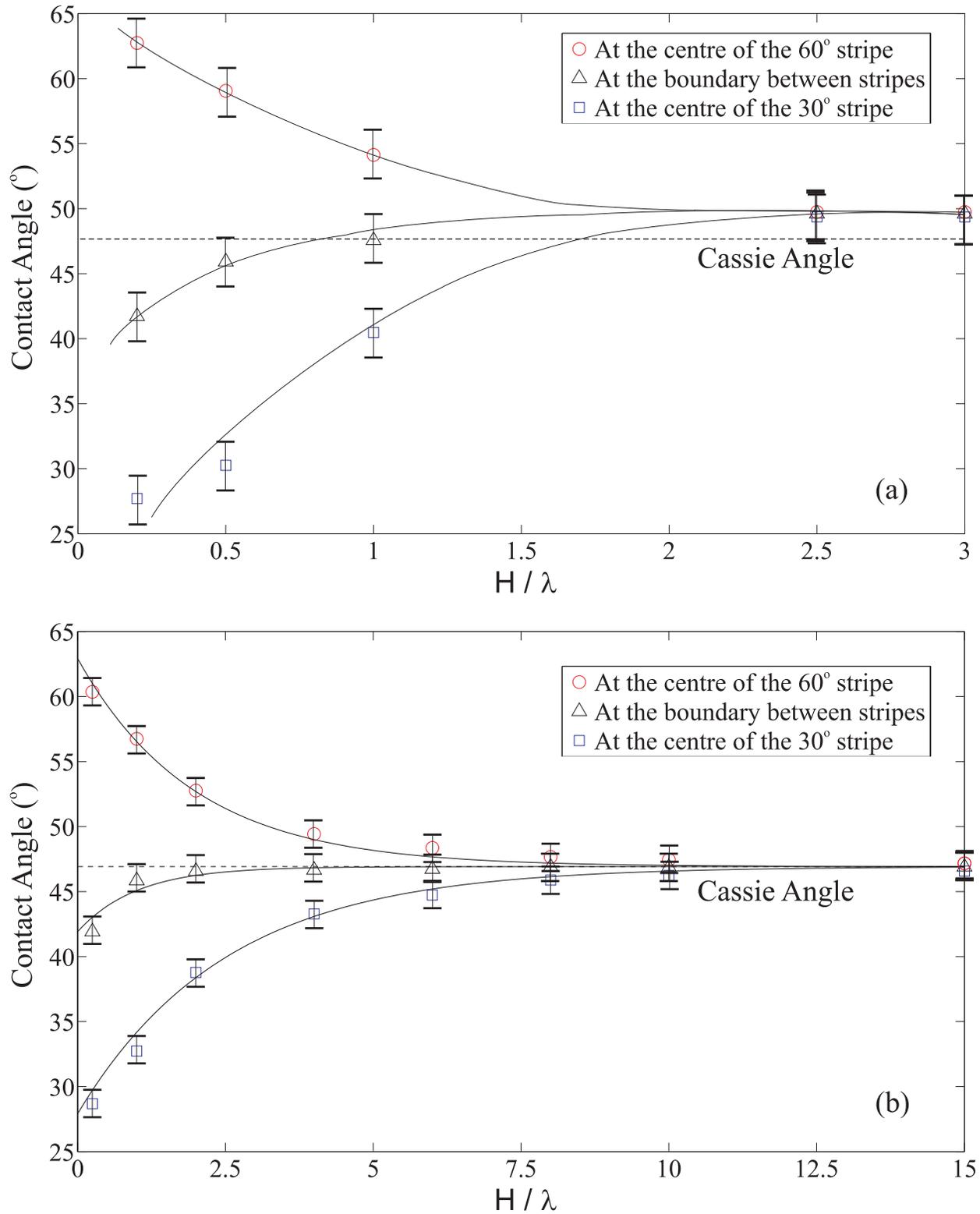}
\caption{Macroscopic contact angle at the centre of the $\theta_i=60^\circ$ stripe, the centre of the 
$\theta_i=30^\circ$ stripe and the boundary between stripes as a function of the height of the liquid bridge $H$. The 
interface thickness is set to (a) $\xi/\lambda=0.016$ (diffuse interface model) and (b) $\xi/\lambda=0.0$ (Surface 
Evolver). The error bars in (a) correspond to $3^\circ$ and in (b) $2^\circ$. The lines are a guide to the eye.}
\label{angles}
\end{center}
\end{figure}

The variation of the macroscopic contact angles for different values of $H/\lambda$, shown in Figure \ref{angles}, 
are a consequence of the behaviour described above. Figures \ref{angles}(a) and (b) were obtained using the diffuse 
interface model ($\xi/\lambda=0.016$) and Surface Evolver (without line tension) respectively. For large $H/\lambda$, 
the macroscopic contact angle is independent of the position across the pattern, i.e. the contact angle multiplicity 
is mitigated. The value of the macroscopic contact angle is consistent with the Cassie angle, $\theta_{C}$, which 
corresponds to the arccosine of an average of the cosines of the intrinsic contact angles of the stripes. 
\cite{cassie} The macroscopic contact angle obtained from the diffuse interface model deviates slightly from the 
Cassie angle (by $\sim2^\circ$). This deviation may be due to line tension effects and/or uncertainties in the 
simulation method. Drelich et. al. \cite{Drelich1993379} pointed out that, when the contact line is contorted, there 
is a correction to the value of the effective contact angle due to the line tension. Simple estimates from our 
simulations show that the correction is of order $1^\circ$. Uncertainties in the lattice Boltzmann simulations arise 
from the discretisation errors in the implementation and measurement of the contact angle. This uncertainty is of 
order $2-3^\circ$. For the Surface Evolver data, the uncertainty comes from the measurement of the contact angle and is
typically of order $1-2^\circ$. For small $H/\lambda$, the macroscopic contact angle at the center of the stripes mirrors the local contact angle, 
while at the boundaries, it takes an intermediate value between the two local contact angles.

\begin{figure} 
\begin{center}
\includegraphics[width=\columnwidth,angle=0]{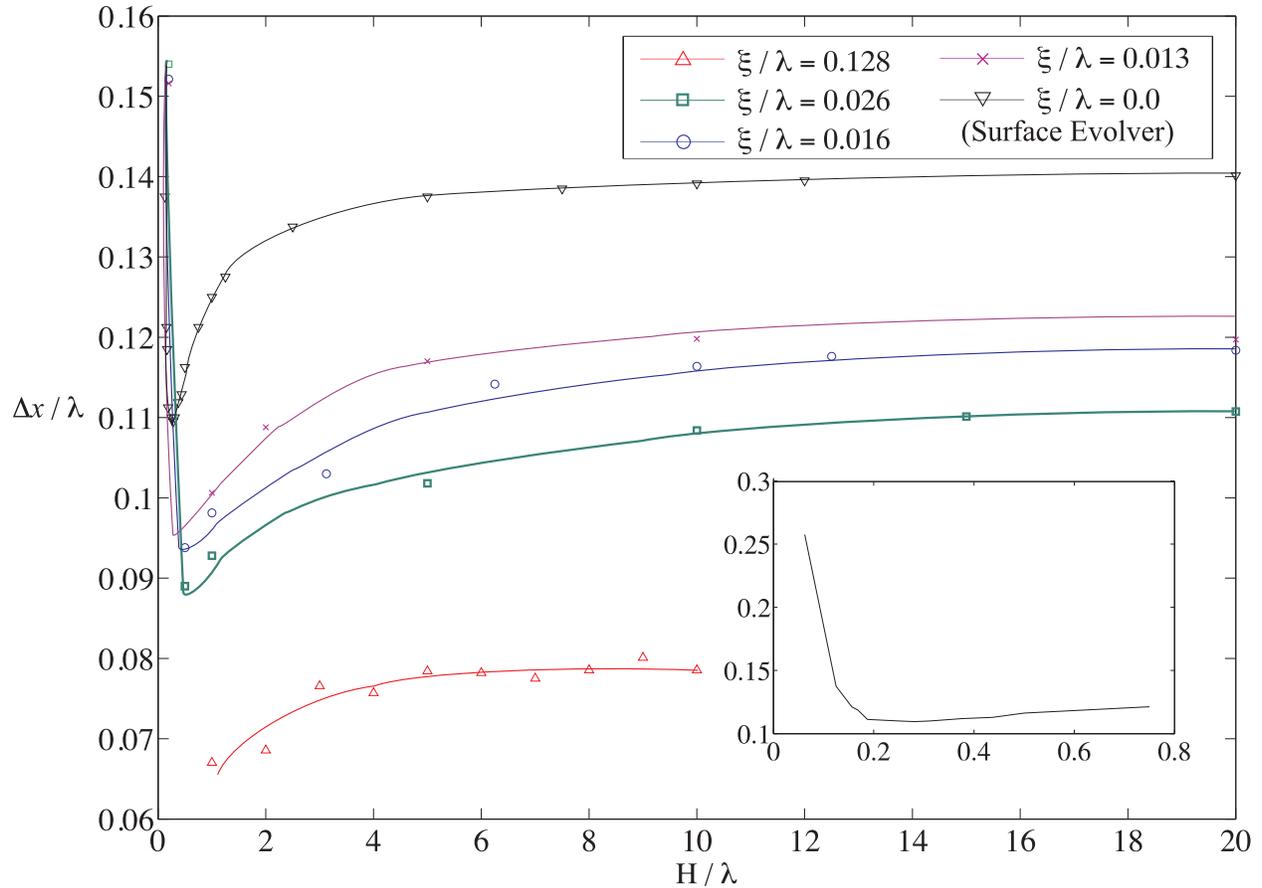}
\caption{Lattice Boltzmann results for the variation of the magnitude of the contact line corrugation $\Delta x$ with the height of the liquid bridge 
$H$ for different values of interface thickness $\xi$. All lengths are scaled to the spatial period $\lambda$ and the 
lines are a guide to the eye. The inset shows the Surface Evolver data ($\xi/\lambda$=0.0) close to the crossover 
region.}
\label{height}
\end{center}
\end{figure}

Quantitative results showing the crossover between the two regimes described above are shown in Figure \ref{height} 
which presents data for several different values of the reduced interface thickness, $\xi/\lambda$. The $\xi/\lambda 
= 0$ results were obtained using Surface Evolver; the rest are from lattice Boltzmann simulations. In each case we 
observe a similar dependence of corrugation on the distance between the plates. When the height of the channel is 
much larger than the healing length of the interface corrugation, the dimensionless deformation of the contact line 
saturates. This is as predicted by the classical theory of capillarity, because the contact line corrugation must be 
proportional to the characteristic length of the pattern \cite{neumann,Degennes}. As $H/\lambda$ is decreased there 
is a slight decrease in $\Delta x/ \lambda$ because the larger Laplace curvature between the plates inhibits 
corrugation. The typical value of $\Delta x/ \lambda$ obtained here is consistent with previous work by Hoorfar {\it 
et. al.} \cite{Hoorfar}. 

As $H/\lambda$ is decreased below 1, the distortion first decreases and then increases sharply. This increase 
occurs because the interface is in the regime where it remains corrugated for all values of $z$ and smaller $H$ 
reduces the excess interface free energy resulting from the corrugation, but not the wetting energy gained at the 
plates. In the inset of Figure \ref{height} we show the Surface Evolver data close to the crossover region.

\begin{figure} 
\begin{center}
\includegraphics[width=\columnwidth,angle=0]{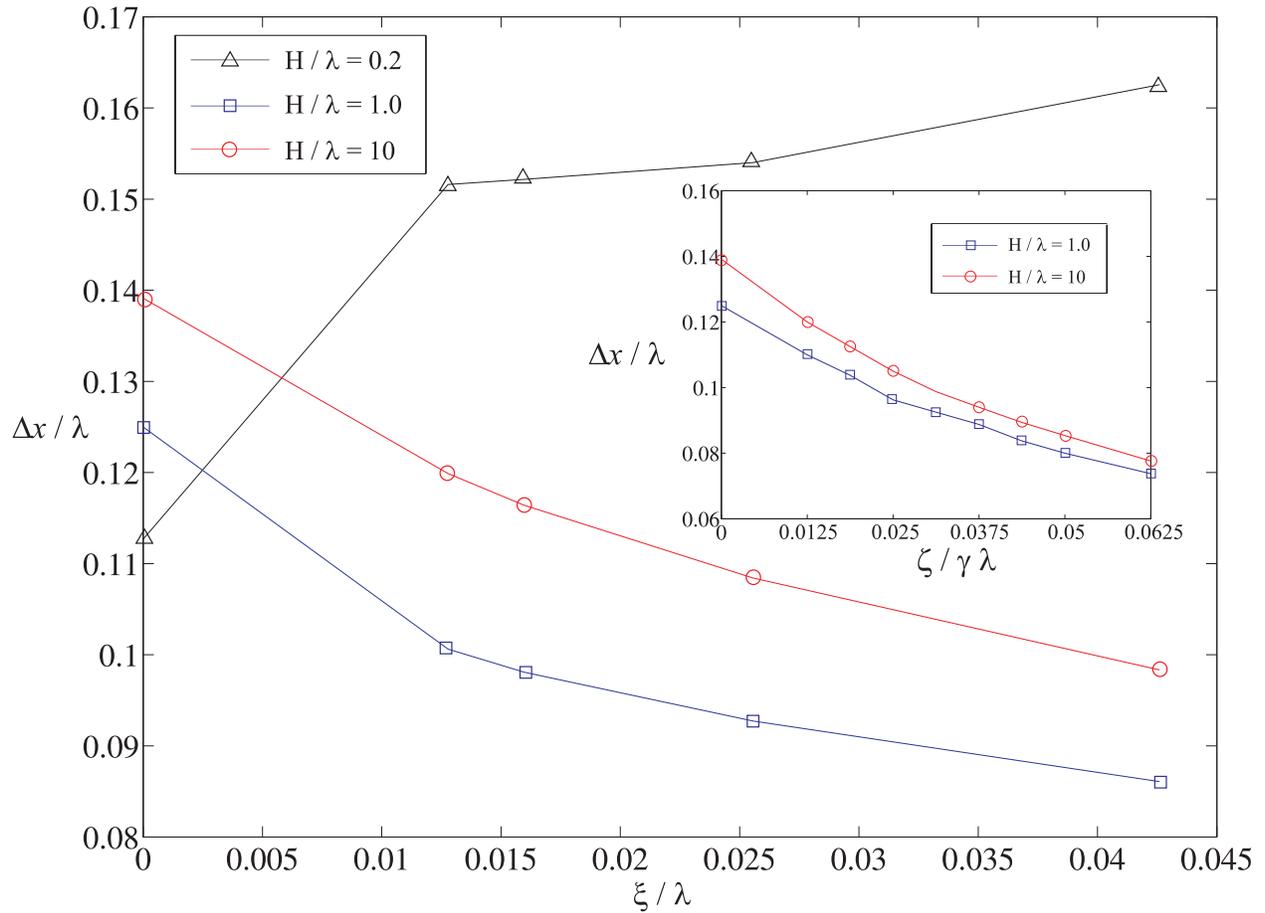}
\caption{Variation of the magnitude of the contact line corrugation $\Delta x$ with the thickness of the liquid--gas 
interface $\xi$ for different values of the height of the liquid bridge $H$.  All lengths are scaled to the spatial 
period $\lambda$. The inset shows how the corrugation depends on the magnitude of the line tension $\zeta$ in Surface 
Evolver. The line tension is scaled to the liquid--gas surface tension $\gamma$ and the spatial period $\lambda$. 
$\lambda$ is set to 0.8 and the lines are a guide to the eye.}
\label{width}
\end{center}
\end{figure}

To explore the variation of the corrugation with interface thickness more closely, $\Delta x/ \lambda$ is plotted 
against $\xi/\lambda$ for three different values of $H/\lambda$ in Figure \ref{width}. For $H/\lambda=1$ and $10$ the 
contact line distortion decreases slightly as the interface thickness becomes larger, reminiscent of the 
flattening-out effect that would result from including a line tension in the free energy. This suggests that the 
lattice Boltzmann model incorporates an effective line tension, due to discretisation effects, which increases slowly 
with increasing interface thickness. In the inset of Figure \ref{width}, we plot $\Delta x/ \lambda$ against 
$\zeta/(\gamma\lambda)$ obtained using Surface Evolver. $\zeta$ and $\gamma$ are defined as the line tension and the 
liquid--gas surface tension respectively. The Surface Evolver results show that $\Delta x/ \lambda$ has a similar 
dependence on $\zeta/\gamma$, as on $\xi$ for the lattice Boltzmann simulations. This dependence of $\Delta x/ 
\lambda$ on the line tension is also consistent with previous studies by Neumann {\it et. al.} \cite{Amirfazli} (and 
the references therein). For $H/\lambda=0.2$, however, the distortion increases slightly as the interface thickness 
becomes larger indicating that a diffuse interface favours corrugations along the solid surface. 
This behaviour is not reproduced in Surface Evolver when we take into account the effect of positive line tension.

\section{Discussion}
\label{discussion}

We have presented numerical results for the behaviour of the interfaces bounding a liquid bridge confined between two 
plates patterned by stripes of differing contact angle for the particular case where the contact line runs, on 
average, perpendicular to the stripes. We were able to see clearly, using both a diffuse interface approach and 
Surface Evolver, a sharp crossover between a regime where the interface corrugations on the two surfaces are 
independent of each other, and decay moving away from the substrates, to a regime where the corrugation persists 
across the bridge. In the former case it is possible to define a unique macroscopic contact angle for the drop, as 
predicted by the Cassie equation. In the second the macroscopic contact angle varies between the intrinsic values on 
different areas of the surface, as predicted by the local Young equation. The crossover occurs for $H/\lambda$ of 
order unity. In both regimes, the interface diffuseness plays a relevant role through the interface thickness, which 
is related to the line tension. To compare the simulation parameters we have considered here to the physical 
variables, we note that the typical value of $\zeta/(\gamma\lambda)$ in our paper is $10^{-2}$. Using $\gamma = 
10^{-2}$ J/m$^2$ and $\lambda$ = 100 nm$-$1 $\mu$m, this corresponds to $\zeta$ = $10^{-11}-10^{-10}$ J/m. This value 
of line tension is comparable to those reported in experiments \cite{Pompe}.

Results from the diffuse interface algorithm approach those obtained using Surface Evolver (with line tension) in the 
limit that the interface thickness goes to zero. 
The advantages of Surface Evolver are that it is considerably quicker (typically by one or two orders of magnitude), 
that the line tension can be controlled, and that it immediately accesses the physical limit of an interface which is 
sharp on micron length scales. The advantages of diffuse interface models, on the other hand, is that they can model 
drop hydrodynamics, that they can follow changes in the topology of the liquid, such as drop break up, and that they 
can model pinning and depinning correctly beyond the quasistatic limit. It is pleasing that, for a problem such as 
this where both approaches should be applicable, the results for drop shapes are comparable.

\acknowledgement

This work was supported by the "Ministerio Espa\~nol de Educaci\'{o}n y Ciencia" (project
MAT2007-66117 and contract "Ram\'{o}n y Cajal" RYC-2005-000983), Junta de Andaluc\'{i}a (project P07-FQM-02517), the 
European Social Fund (ESF) and the EU project INFLUS.


\bibliography{references}

\end{document}